\documentclass{amsart}
 \usepackage[foot]{amsaddr}

\usepackage[inner=2cm,outer=2cm, top=3cm, bottom=3cm]{geometry}

\usepackage{amssymb}
\usepackage{graphicx}
\usepackage{tabularx}
\usepackage{subfigure}

\title{Equation of state and optical properties\\
of shock-compressed C:H:N:O molecular mixtures} 

\author{M. Guarguaglini$^{1, 2}$}

\author{J.-A. Hernandez$^{1, 2}$}

\author{T. Okuchi$^{3}$}

\author{P. Barroso$^{4}$}
\author{A. Benuzzi-Mounaix$^{1, 2}$}
\author{R. Bolis$^{1, 2}$}
\author{E. Brambrink$^{1, 2}$}
\author{Y. Fujimoto$^{5}$}
\author{R. Kodama$^{5, 6, 7}$}
 \author{M. Koenig$^{1, 2, 6}$}
\author{F. Lefevre$^{1}$}
\author{K. Miyanishi$^{7}$}
\author{N. Ozaki$^{5, 7}$}
\author{T.Sano$^{7}$}
\author{Y. Umeda$^{5}$}
\author{T. Vinci$^{1, 2}$}
 
\author{A. Ravasio$^{1, 2}$}

\email{marco.guarguaglini@polytechnique.edu}

\address{$^{1}$LULI - CNRS, \'Ecole Polytechnique, CEA, Universit\'e Paris-Saclay, route de Saclay, 91128 Palaiseau cedex, France}
\address{$^{2}$Sorbonne Universit\'e, UPMC Univ. Paris 06, CNRS, Laboratoire d'Utilisation des Lasers Intenses (LULI), place Jussieu, 75252 Paris cedex 05, France}

\address{$^{3}$Institute for Planetary Materials, Okayama University, Misasa, Tottori 682-0193, Japan}
\address{$^4$GEPI, Observatoire de Paris, PSL Universit\'e, CNRS, 77 avenue Denfert Rochereau, 75014 Paris, France}
\address{$^5$Graduate School of Engineering, Osaka University, Suita, Osaka 565-0871, Japan}
\address{$^6$Open and Transdisciplinary Research Initiatives, Osaka University, Suita, Osaka 565-0871, Japan}
\address{$^7$Institute of Laser Engineering, Osaka University, Suita, Osaka 565-0871, Japan}
\begin{document}

\begin{abstract}
Water, ethanol, and ammonia are the key components of the mantles of Uranus and Neptune. To improve structure and evolution models and give an explanation of the magnetic fields and luminosities of the icy giants, those components need to be characterised at planetary conditions (some Mbar and a few $10^3$ K). Those conditions are typical of the Warm Dense Matter regime, which exhibits a rich phase diagram, with the coexistence of many states of matter and a large variety of chemical processes. H$_2$O, C:H:O, and C:H:N:O mixtures have been compressed up to 2.8 Mbar along the principal Hugoniot using laser-driven decaying shocks. The experiments were performed at the GEKKO XII and LULI 2000 laser facilities using standard optical diagnostics (Doppler velocimetry and pyrometry) to characterise equation of state and optical reflectivity of the shocked states. The results show that H$_2$O and the C:H:N:O mixture share the same equation of state with a density scaling, while the reflectivity behaves differently by what concerns both the onset pressures and the saturation values. The reflectivity measurement at two frequencies allows to estimate the conductivity and the complex refractive index using a Drude model.
\end{abstract}

\flushbottom
\maketitle
%
%
\thispagestyle{empty}
\pagestyle{plain}


\section*{Introduction}

Composite mixtures behaviour at extreme pressures and temperatures shows intriguing chemical and physical processes, involving complex bounding scenarios. At different temperatures and densities, a variety of states exist. These include combinations of many chemical species in distinct states of ions, atoms, molecules, clusters and lattices, depending on the specific conditions. Of particular interest are C:H:N:O mixtures (also called planetary ices), as they comprise the major component of the interiors of our icy giant planets Uranus and Neptune. Their structure are indeed composed by an outer layer of hydrogen and helium, an ``icy'' mantle made of a water (H$_2$O) - methane (CH$_4$) - ammonia (NH$_3$) mixture, and possibly a rocky core \cite{guillot99}. As pressure and temperature increase from the outer layers towards the core, their interiors are expected to exhibit a wide range of different states embracing atomic and molecular fluids, dissociated plasmas, and superionic lattices.\\
The complexity in describing the behaviour of these mixtures at planetary conditions (few Mbar, few $1000$ K) is at the basis of the numerous lacunae in our understanding of Uranus and Neptune. Their internal structures are inferred from the observed gravitational fields, masses, internal rotation and radii. However mass distribution remains ambiguous \cite{podolak12}. Accurate analysis of Voyager 2 data \cite{lindal87, anderson87, tyler89, lindal92} even open the possibility for a dichotomy in their structures, indicating that the two planets could have very different interiors, despite being similar in mass and radius. Lack of precise information on transport properties of the C:H:N:O mixture is also casting serious issues in explaining Uranus and Neptune's magnetic fields \cite{stevenson10}. Similarly, the simplified approach adopted in the ice characterisation fails in describing Uranus' low luminosity \cite{pearl90, pearl91}. Resolving this situation is even more urgent today as the discovery of exoplanets is incredibly active. Since solar planets are used as prototypes for extrasolar planets \cite{swift12}, the loose description of planetary ice and the resulting approximate portrait of Uranus and Neptune not only prevent the understanding of extrasolar giant planets such as GJ 436b or HAT-P-11b but also affect our capability to distinguish Earth-like planets candidates. As a result there is actually a great need to establish benchmarking values for the equations of states, phase diagrams, and transport properties of H$_2$O-NH$_3$-CH$_4$ mixtures at Mbar pressures and temperatures of some $10^3$ K. So far, our knowledge of water/methane/ammonia mixtures mainly relies on ab initio calculations \cite{chau11, lee11, meyer15, bethkenhagen17} since experimental data at planetary conditions \cite{radousky90, nellis97, chau11} are limited, while water has been experimentally probed up to high pressures \cite{mitchell82, lyzenga82, knudson12, kimura15, millot18}.\\

In the present work, we have compressed water and two C:H:(N):O mixtures relevant for ice giant interiors up to $2.8$ Mbar through laser-driven shocks. We have measured the equation of state of the shocked sample and the optical reflectivity of the shock front using optical diagnostics (VISARs and SOP). An estimation of the electronic contribution to conductivity is given using a Drude model.\\

\section*{Methods and experimental setup}

\subsection*{Mixtures.} Liquid water/ethanol (C:H:O) and water/ethanol/ammonia (C:H:N:O) mixtures have been prepared by adding up different amounts (see the Supporting Information) of pure water, pure ethanol, and a liquid water/ammonia (28\% wt.) mixture to obtain the following atomic ratios: C:H:O = 4:22:7; C:H:N:O = 4:25:1:7. The latter - \textit{synthetic Uranus} \cite{nellis97} - reproduces the chemical composition of Uranus and Neptune's mantles, with the C:N:O abundance ratios comparable to those of the Solar System \cite{cameron73}. The density of the mixtures at ambient conditions was $\rho_0^{mix} = 0.885$ g/cm$^3$.\\
\subsection*{Laser facilities.} Experiments were performed at the GEKKO XII laser facility of the Institute of Laser Engineering, Osaka University (Japan) \cite{ozaki04} and at the LULI 2000 laser facility of the Laboratoire d'Utilisation des Lasers Intenses, \'Ecole Polytechnique (France). At GEKKO XII, we used 3 up to 9 beams (corresponding to  energies on target from 120 - 440 J) at 351 nm, with a $600$ $\mu$m focal spot diameter. At LULI 2000 we used 1 or 2 beams (energies on target from 200 - 500 J) at 527 nm, with a $500$ $\mu$m focal spot diameter. In both cases, the laser pulse duration was $2.5$ ns and phase plates were used to obtain a uniform irradiation spot.\\
\subsection*{Targets.} To optimise the target design and ensure there were no shock reverberations in the sample, we simulated laser-target interaction and the shock loading into the cell with the Lagrangian 1-D hydrodynamic code MULTI \cite{ramis88}. The equation of state of the target components were extracted from the SESAME tables \cite{lyon92, johnson94}. The table 7154 for water has been used for the mixtures. The multi-layered target cells were composed by a 10-15 $\mu$m thick CH ablator, a 40  $\mu$m thick Al shield, a 50 $\mu$m thick $\alpha-$SiO$_2$ standard, the sample (4 mm thick), and a rear $\alpha-$SiO$_2$ window (200 $\mu$m thick). We also performed some high-intensity shots at GEKKO XII with CH 50 $\mu$m / Au 3  $\mu$m / Al 5  $\mu$m / $\alpha-$SiO$_2$ 20 $\mu$m / mixture 4 mm / $\alpha-$SiO$_2$ 200 $\mu$m targets, the gold layer serving as X-ray shield to prevent any pre-heating of the sample. As probe lasers, we used a YAG at 532 nm at GEKKO XII and at 532 and 1064 nm at LULI 2000, with a full-width half-maximum pulse duration of $\sim 10$ ns.\\

\begin{figure}[h!]
\centering
\includegraphics[width=0.6\columnwidth]{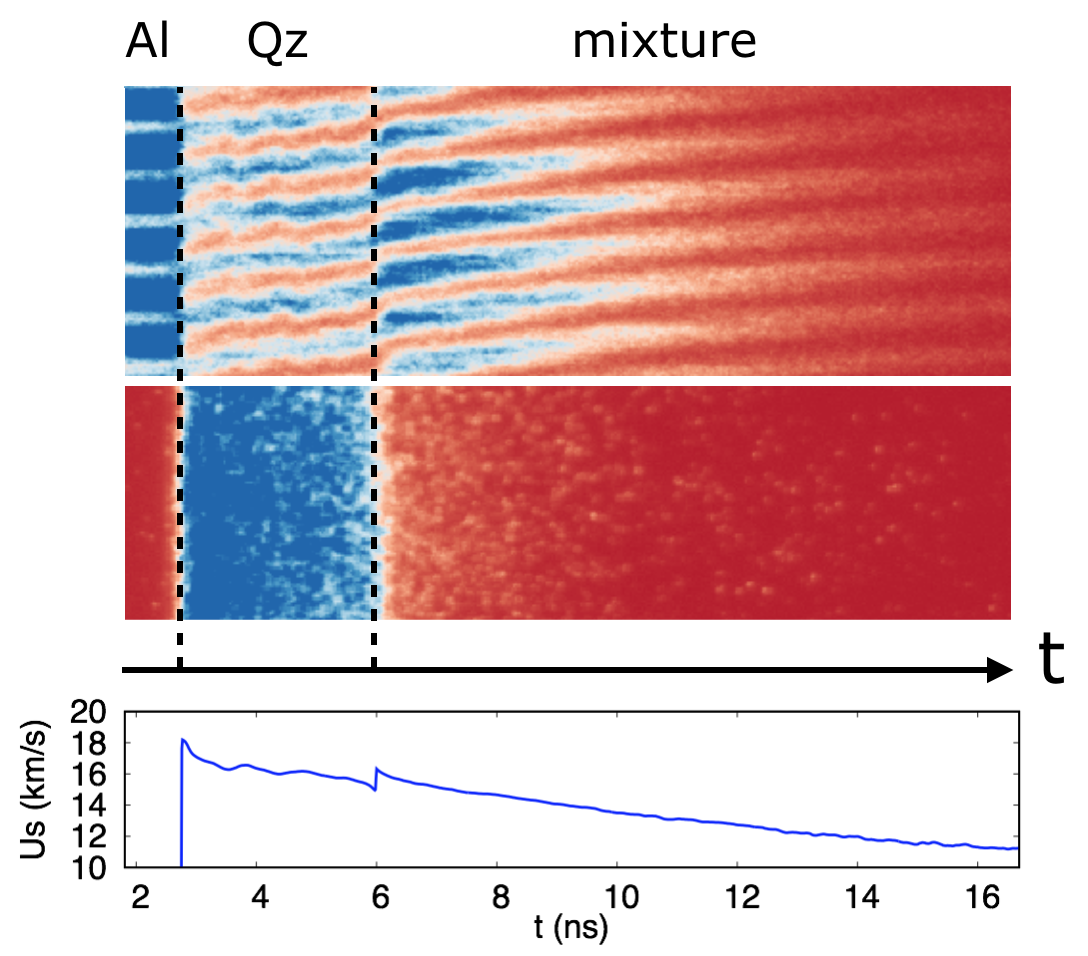}
\caption{\textbf{Top.} VISAR and SOP raw output for a GEKKO shot on C:H:N:O mixture. The three time periods indicate when the probe laser is reflected by aluminum (Al), when a reflecting shock front is propagating through the quartz layer (Qz) and the mixture. The transverse target dimension is $\sim 180$ $\mu$m. \textbf{Bottom.} Shock velocity temporal profile from the VISAR raw output.}
\label{fig:diag_output}
\end{figure}

\subsection*{Data analysis.} Time-resolved shock velocity $U_s (t)$ has been extracted using the \textit{Neutrino} software \cite{flacco11} from the output of two VISARs \cite{dolan06} (Doppler velocity interferometers, see the Supporting Information), both working at 532 nm (GEKKO XII) or one at 532 and one at 1064 nm (LULI 2000). The thermodynamic conditions (the mass density $\rho$, the pressure $p$, and the internal energy density $E$) reached in the mixture have been obtained from the Rankine-Hugoniot relations \cite{rankine70, hugoniot87} through impedance mismatching \cite{forbes}, using quartz as \textit{in situ} standard. To span a range of thermodynamical conditions with a single shot, we employed a decaying shock technique. We determined the shock velocity at the exit from quartz and at the entrance in the mixture ($U_s^{Qz}$ and $U_s^{mix}$, respectively) with a linear fit on $U_s (t)$ on a time window of some ns before and after the crossing of the quartz/mixture interface. Shock velocity \textit{versus} fluid downstream velocity ($U_s$-$U_p$) Sandia Z-pinch data \cite{knudson09} have been used as reference for quartz. The adiabatic release of quartz on the lower-impedance mixture at the shock crossing of the interface has been modeled using a mirror Hugoniot approximation. This method agrees with the use of a quartz release model \cite{knudson13} within the error bars in the region where the latter can be applied.\\
Time-resolved self emission has been measured through a streaked optical pyrometer (SOP) working at $\lambda_{SOP} = 455$ nm. Temperature has been obtained from Planck's law $T = T_0 / \ln(1 + A  \epsilon_{\lambda_{SOP}} / N_c)$, where $T_0 = hc / k_B \lambda_{SOP}$, $A$ is a calibration factor, $\epsilon_{\lambda_{SOP}}$ is the emissivity of the shock front at the working wavelength and $N_c$ is the number of counts on the SOP. To get the emissivity at $455$ nm we used the reflectivity measured at $532$ nm under a grey-body hypothesis: $\epsilon_{\lambda_{SOP}} = 1 - R(532 \ nm)$. SOP calibration has been made \textit{in situ}, by determining the $A$ factor using quartz as standard (GEKKO XII) or using a calibration lamp with known emission temperature (LULI 2000, see the Supporting Information).\\
A typical VISAR and SOP output is shown in Figure \ref{fig:diag_output}, together with the extracted shock velocity temporal profile.\\
Since the shocked sample has different thermodynamical conditions with respect to the un-shocked one, its refractive index changes. According to the Fresnel equations, the shock front reflects a fraction of the incident light from the probe laser. The reflectivity of the shock front has been measured with the VISARs as the ratio between the shot signal and a reference signal reflected on the aluminum/quartz interface. A VISAR-independent reflectivity measurement at $532$ nm was included in the setup for some shots at LULI 2000. Since VISARs and reflectometer can measure only a relative value, a quartz reflectivity fit \cite{millot15} based on previous measurements \cite{hicks06}, has been used to calibrate the measure.\\

\section*{Results and discussion}

\subsection*{Equation of state}

\begin{figure}[h!]
\centering
\includegraphics[width=\textwidth]{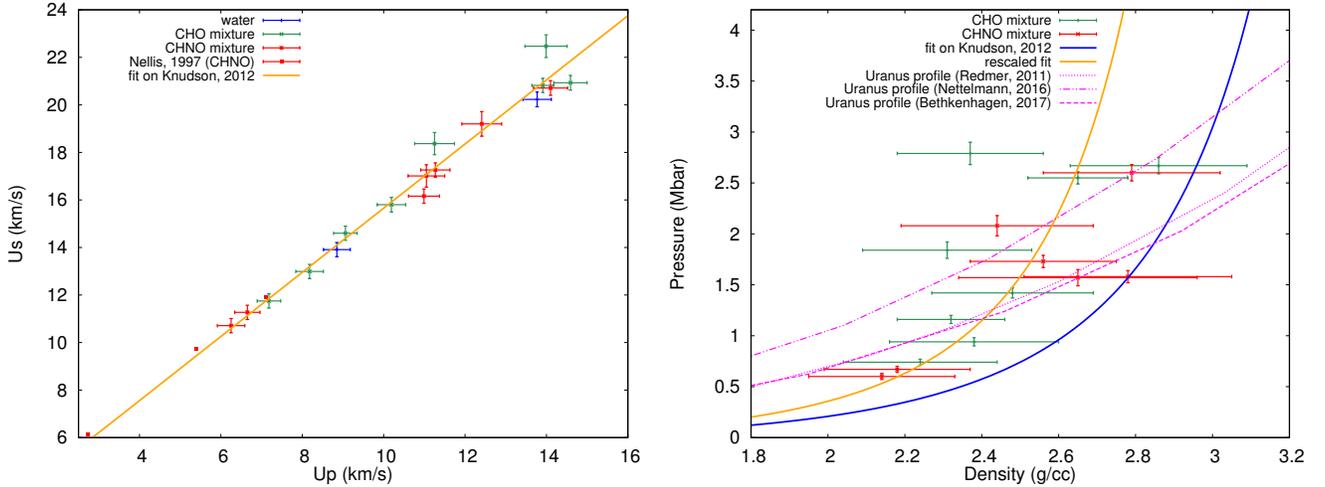}
\caption{\textbf{Left.} Water, C:H:O, and C:H:N:O mixture $U_s - U_p$ relation along the principal Hugoniot with a fit on previous results \cite{knudson12}. \textbf{Right.} C:H:O and C:H:N:O mixture $p - \rho$ relation along the principal Hugoniot. The blue line is the $p - \rho$ transposition of the $U_s - U_p$ linear fit on previous water data \cite{knudson12}. The orange line is the same fit rescaled to take into account the initial density difference between water and mixtures. The magenta lines are Uranus profiles according to water-only model \cite{redmer11},  a model with a thermal boundary layer \cite{nettelmann17}, and an icy model \cite{bethkenhagen17}.}
\label{fig:mixtures_usup_rhoP}

\end{figure}

Figure \ref{fig:mixtures_usup_rhoP} (left) shows $U_s$-$U_p$ velocity data for pure water and the C:H:O and C:H:N:O mixtures, together with the best linear fit on previous water high pressure shock data \cite{knudson12}: $U_s =1.35 U_p + 2.16$ km/s. We extracted $\rho$-$p$-$E$ thermodynamic conditions from the $U_s$-$U_p$ relation via the Rankine-Hugoniot equations:

\begin{subequations}
\begin{align}
\rho & = \rho_0 \frac{U_s}{U_s - U_p} \\
p & = p_0 + \rho_0 U_s U_p \\
E  & = E_0 + \frac{1}{2} \left(p + p_0 \right) \left(\frac{1}{\rho_0} - \frac{1}{\rho}\right).
\end{align}
\end{subequations}

$\rho$-$P$-$E$ results are shown in Table \ref{tbl:exp_data}. We observed that the $U_s$-$U_p$ relation does not significantly change between water and C:H:(N):O mixture. Therefore, the only discrepancy in their $p$-$\rho$ relation along the principal Hugoniot is due to the different initial density (1.00 \textit{vs} 0.88 g/cm$^3$). The mixtures $p$-$\rho$ relation is shown in Figure \ref{fig:mixtures_usup_rhoP} (right) and compared with a fit on previous water data \cite{knudson12}. This fit has been rescaled to take into account the different initial density of water and mixtures in order to be immediately compared with mixture data.\\
The common $U_s$-$U_p$ relation between water and mixtures indicates that they share a similar structural behaviour. This confirms previous first-principles molecular dynamics (FPMD) simulations \cite{chau11} which identify the regime we explored as an electronic conducting phase. At these temperatures, carbon-carbon and carbon-nitrogen bond lifetimes are predicted to be very short by first-principles calculations \cite{meyer15}. This prevents polymerisation and clustering, and causes the existence of an atomic fluid above $5000 - 6000$ K. Therefore, no structural effect of carbon and nitrogen atoms on the $U_s$-$U_p$ relation is expected at those conditions. These results confirm recent FPMD calculations \cite{bethkenhagen17} which validate the use of the linear mixing approximation when dealing with C:H:N:O mixtures at planetary conditions.\\

%
%
%

\begin{figure}[h!]
\centering
\includegraphics[width=0.7\textwidth]{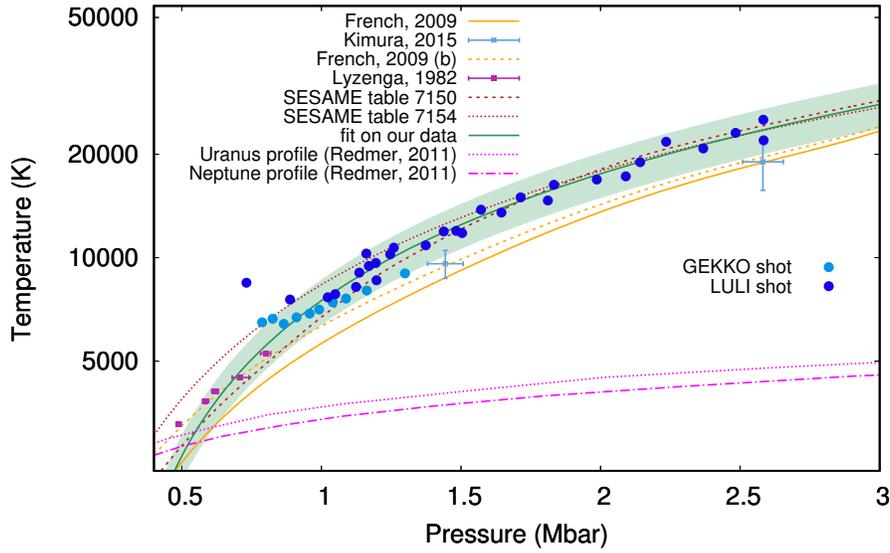}
\caption{Water temperature \textit{vs} pressure along the principal Hugoniot. Each color dot is a decaying shock measurement. The green-shaded area corresponds to the fit within the errors. Uranus and Neptune isentropes are from a water-only planetary model \cite{redmer11}.}
\label{fig:water_T}
\end{figure}

\begin{figure}[h!]
\centering
\includegraphics[width=0.7\textwidth]{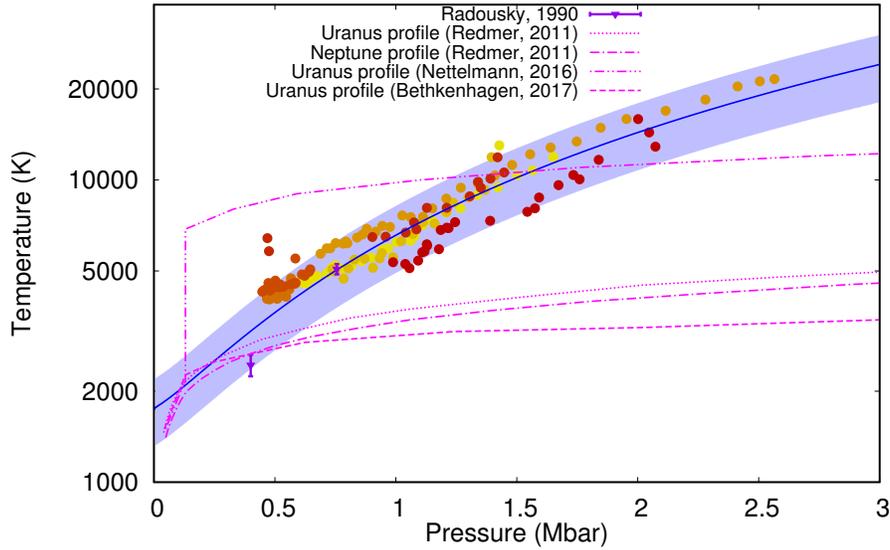}
\caption{C:H:N:O mixture temperature \textit{vs} pressure along the principal Hugoniot. Color dots are decaying shock measurements, each color corresponds to a different shot. The blue-shaded area corresponds to the fit within the errors. Violet points with error bars are previous data along the principal Hugoniot \cite{radousky90}. A water-only planetary model \cite{redmer11} is shown for Uranus and Neptune. A thermal boundary layer \cite{nettelmann17} and an icy \cite{bethkenhagen17} model are shown for Uranus.}
\label{fig:CHNO_T}
\end{figure}

The temperature-pressure ($T$-$p$) relations of water and C:H:N:O mixture are shown in Figure \ref{fig:water_T} and \ref{fig:CHNO_T}, respectively. Figures \ref{fig:water_T} and \ref{fig:CHNO_T} also show the predicted planetary isentropes of Uranus \cite{redmer11, nettelmann17, bethkenhagen17} and Neptune \cite{redmer11}. Our data have been fitted with the function $T (U_s) = \theta_0 + \gamma  U_s^{\delta}$, rescaled to be pressure-dependent using the $p(U_s)$ relation given by the equation of state of water. 
An extrapolation of our fit to lower pressures is compatible with previous gas-gun data \cite{lyzenga82}. While our data agree within the errors with recent laser shock results \cite{kimura15}, our temperatures are higher than those given by FPMD simulations \cite{french09}, although it is worth noticing that when quantum corrections from molecular vibrations are taken into account \cite{french09b} the predicted temperatures increase of $\simeq 700$ K and become more similar to our data.\\
Our $T$-$p$ results for C:H:N:O agree with a previous low-pressure experimental study of the same mixture \cite{radousky90}. Water and C:H:N:O temperatures on the principal Hugoniot are comparable, although the relatively high error bars on temperature make difficult to point out possible discrepancies.\\
Hereafter we compare our $T$-$p$ data with the available models of planetary interiors profiles. Most of them \cite{redmer11, bethkenhagen17} predict an adiabatic profile inside the icy giants, implying that temperatures stay relatively low (3 - 4 $\cdot 10^3$ K) even at the highest pressures we explored (about 3 Mbar). Indeed, we compressed the sample through single shock loading, which is a process causing high entropy increase, reaching higher temperatures than those of isentropic models. Nevertheless, recent interior profile models include a thermal boundary layer \cite{nettelmann17}, predicting 2-3 times higher temperatures, consistent with our data up to $1.5$ Mbar. Moreover, the recent discovery of a large amount of exoplanets exhibits a wide range of structures including hot Neptunes, whose interior profiles can match the thermodynamical conditions we explored. Finally, our data are useful for the validation of FPMD simulations at extreme conditions.\\

\subsection*{Optical reflectivity}

\begin{figure}[h!]
\centering
\includegraphics[width=\textwidth]{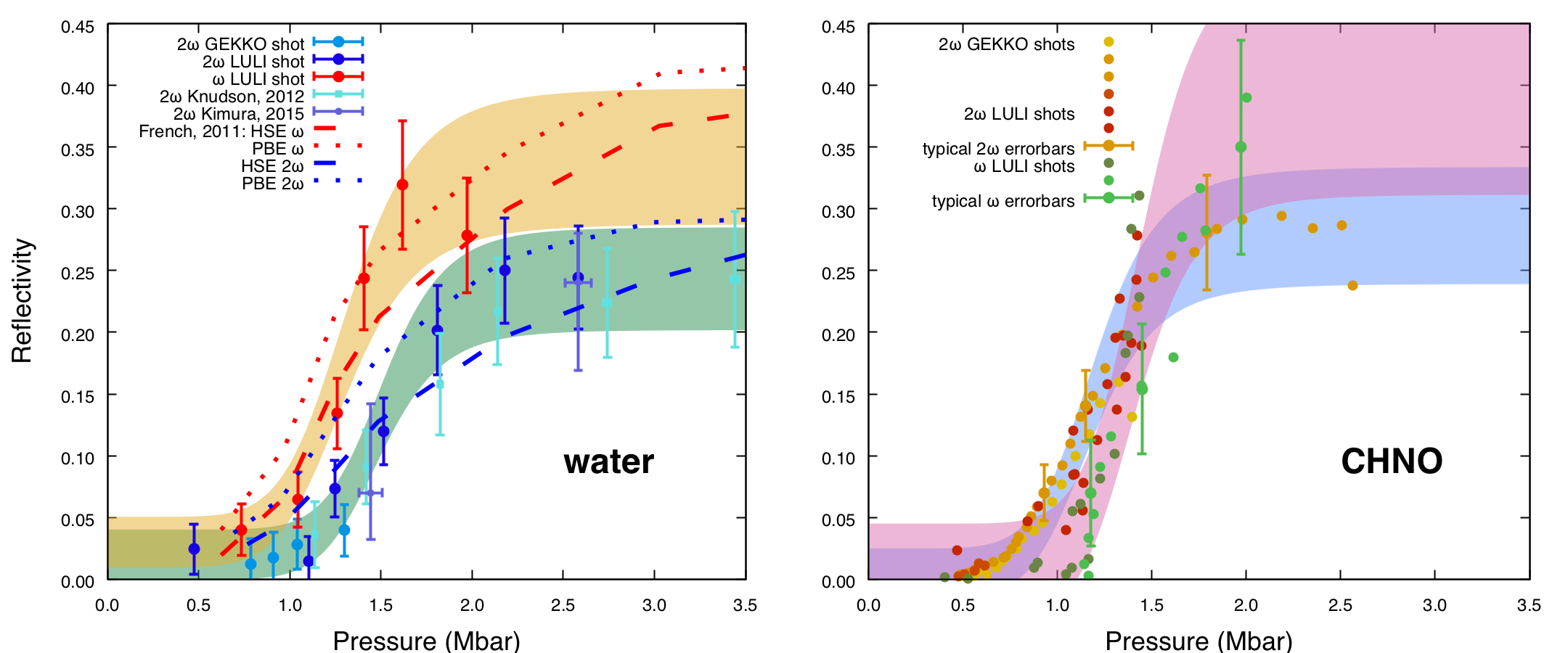}
\caption{\textbf{Left.} Water reflectivity at $532$ and $1064$ nm \textit{vs} pressure along the principal Hugoniot. Color dots are decaying shock measurements, each color corresponds to a different shot. Green and gold-shaded areas correspond to the fit on our data at $532$ and $1064$ nm within the error bars, respectively. Dashed and dotted lines correspond to DFT / Kubo-Greenwood calculations of the reflectivity \cite{french11} at $1064$ or $532$ nm, respectively, using the HSE (red) or PBE (blue) exchange-correlation functionals. \textbf{Right.} C:H:N:O mixture reflectivity at $532$ and $1064$ nm \textit{vs} pressure along the principal Hugoniot. The blue and pink-shaded area correspond to the fit on our data at $532$ and $1064$ nm within the error bars, respectively. Some typical error bars for reflectivity measurements at $532$ and $1064$ nm have been shown. Error bars at $1064$ nm are larger because of the limited number of available shots.}
\label{fig:R_P}
\end{figure}

The optical reflectivity $R$ of the water shock front at $532$ and $1064$ nm as a function of pressure is shown in Figure \ref{fig:R_P} (left). The gradual increase of reflectivity up to a saturation value along the principal Hugoniot indicates a smooth transition from an insulating to an electronically conducting (``metallic") state with the increase of density, pressure, and temperature. We performed a best fit on each $R$-$U_s$ relation using a Hill function $R(U_s) = R_0 + (R_{sat} - R_0) \cdot U_s^k / (U_s^k + U_0^k)$, which is suitable to model this gradual transition. The function has been then rescaled to be pressure-dependent using an experimental water equation of state \cite{knudson12} to link pressure and shock velocity. The error bars of the fit are discussed in the Supporting Information and mainly depend on the calibration. According to the Hill fit on $532$ nm data, the onset of reflectivity in water occurs at $1.1 - 1.2$ Mbar. At these pressures the reflectivity of the shock front reaches the $10 - 20\%$ of the saturation value, respectively. We found a saturation value of reflectivity of $24 \%$ at $532$ nm and $34 \%$ at $1064$ nm. Our $532$ nm reflectivity data are in quantitative agreement with previous experiments \cite{knudson12, kimura15}. 
When compared with existing calculations \cite{french11}, at $532$ nm, in the low-pressure regime ($P < 1.5$ Mbar), our measured reflectivity is lower than the results using two different exchange-correlation functionals (HSE and PBE). At higher pressures our results are in qualitative agreement with the calculated reflectivity using the PBE functional, while the HSE one fails in providing the correct pressure-dependence, as observed by \cite{knudson12}. At $1064$ nm we always obtained data in qualitative agreement with the calculations.\\
The reflectivity of the C:H:N:O mixture shock front at $532$ nm and $1064$ nm is shown in Figure \ref{fig:R_P} (right) against the shocked sample pressure, with Hill fits on the $R(U_s)$ relations rescaled to be pressure-dependent. These are the first reflectivity data of the C:H:N:O mixture along the principal Hugoniot. There are no calculations of shock-compressed C:H:N:O reflectivity in the literature. The onset of reflectivity in C:H:N:O occurs at lower pressures than in water. Indeed, reflectivity at $532$ nm reaches the $10 - 20\%$ of the saturation value at  $0.8 - 0.9$ Mbar, respectively. While water reflectivity at $1064$ nm is always greater than at $532$ nm, this is not true for C:H:N:O at low pressures: the onset value for the $1064$ nm reflectivity is about $1.2$ Mbar. Starting from $1.5$ Mbar, the $1064$ nm reflectivity becomes greater and saturates at $41\%$. The crossing between the two reflectivity values is an indication of the metallisation of the sample via a gap-closure mechanism. Indeed, frequency-dependent conductivity of a semiconducting state has a maximum at a non-zero frequency. As the gap progressively closes with the increase of density and temperature, conductivity (thus reflectivity) becomes to decrease monotonically with frequency as in a free electron gas \cite{laudernet04, clerouin05, qi15}.\\
The fact that the onset pressure of the C:H:N:O reflectivity at $532$ nm occurs at pressures around $26 \%$ lower than in water can not be fully explained by the $12 \%$ difference between their initial densities. Different dissociation mechanisms that occur in pure water and in carbon-rich mixtures could be at the origin of the mixture higher reflectivity due to the higher free electron density associated to the breaking of carbon-carbon bonds. For similar reasons, the reflectivity saturation value at $532$ nm of C:H:N:O is $29 \%$, higher than that of water ($24 \%$).\\

\subsection*{Conductivity}
\label{sub:cond}

Electrical conductivity is one of the most important parameters to understand the planetary magnetic field generation and structure. Indeed, a dynamo effect can be sustained if magnetic induction dominates over diffusion. This is usually expressed by the requirement that the magnetic Reynolds number $R_m = \mu_0 \sigma u L  \gtrsim 100$ (where $\sigma$ is the electrical conductivity of the active planetary layer component and $u$ and $L$ are the velocity and length scale of the fluid motion inside the layer, respectively).\\
In gas-gun experiments, the DC electrical conductivity can be directly measured using electrodes. This approach can not be applied to laser shock experiments. Instead, they would need a measurement of the complex refractive index of the shocked sample $\tilde{n} = n + i k$ since, from the wave solution of the Maxwell equations, $\sigma(\omega) = 2 \epsilon_0 n(\omega) k(\omega)$. In a restricted range of pressure and temperature the absorption coefficient $\alpha(\omega) = 2 \omega k(\omega) / c$ and the reflectivity 
\begin{equation}
R(\omega) = \frac{\left[ n(\omega) - n_0 (\omega) \right]^2 + k^2 (\omega)}{\left[ n (\omega) + n_0 (\omega) \right]^2 + k^2 (\omega) }
\end{equation}
can be simultaneously measured \cite{millot18}. In this case, the evaluation of the conductivity is straightforward.
Nevertheless, this approach is very delicate and remains restricted to few experiments and conditions. In laser shock experiments only reflectivity is usually measured.
In this case, a model has to be considered in order to infer the complex refractive index. A common approach employs a Drude modelisation of optical properties, modified to account for both free and bound carriers. Even if this model is too simplistic for a well-established conductivity estimation, no first-principles calculations on these mixtures are led to date. This approach can therefore be followed to compare mixture conductivity with water data found in literature and obtained with the same method. 
In the context of the Drude model, $\tilde{n} = \left( n_b^2 + i \sigma(\omega) / \epsilon_0 \omega \right)^{1/2}$, where $n_b$ is the contribution of the bound electrons to the refractive index. $\sigma(\omega) = \sigma_{dc} / (1 - i \omega \tau)$, where $\sigma_{dc}$ is the direct current conductivity and $\tau$ the electron-ion scattering time.\\
The free parameters in this model are $\sigma_{dc}$, $n_b$, and $\tau$. Reflectivity is generally measured at one probe laser frequency (usually in the green: $532$ nm), requiring two of the three parameters to be fixed in an arbitrary way. As $\sigma_{dc}$ is the physical quantity of interest, both $\tau$ and $n_b$ must be estimated. 
A reasonable choice for $\tau$ is the Ioffe-Regel limit $\tau_{IR} = l_s / v_{th}$, which depends on the scattering length $ l_s = 2 \left( 3 M V / 4 \pi N_A N_F \right)^{1/3}$ and on the electron thermal velocity $v_{th} = \left( k_B T / m^\star \right)^{1/2}$. $M$ is the molar mass, $V$ the molar volume, $N_A$ the Avogadro number and $N_F$ the number of atoms in the chemical formula of the mixture; $m^\star = m_e / 2$ is the reduced mass in a semiconductor formalism.
$n_b$ is much more delicate to be estimated over a wide range of pressures and temperatures. It is usually considered either as a constant or linearly dependent on density (Gladstone-Dale model), which is a simplistic assumption for conducting states. The simultaneous measurement of reflectivity at two frequencies in our experiments removes this difficulty, reducing the number of parameters to fix to one ($\tau = \tau_{IR}$).\\
Following this approach, for each Hugoniot state we find the best couple $(n_b, \sigma_{dc})$ which matches the two reflectivity measurements.\\
\begin{figure}[h!]
\centering
\includegraphics[width=0.7\textwidth]{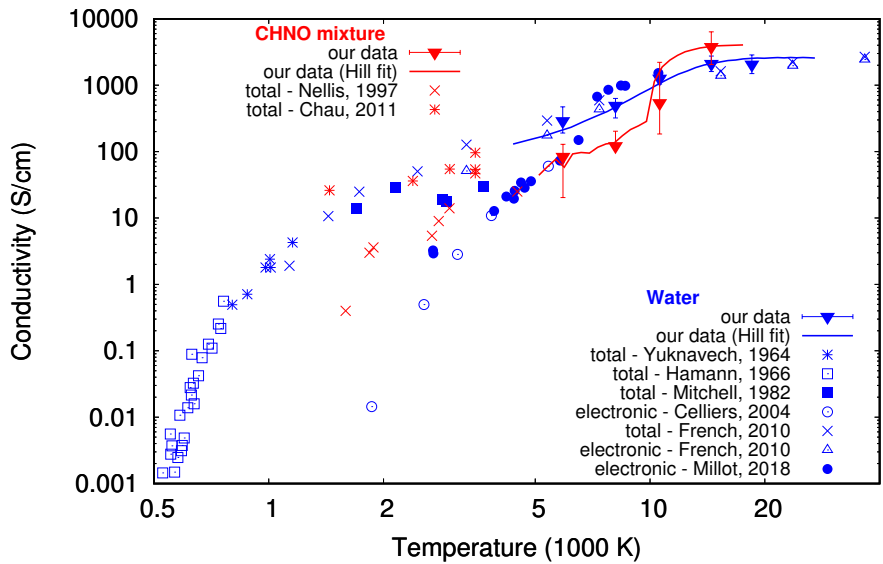}
\caption{Conductivity \textit{vs} temperature for water (blue) and C:H:N:O mixture (red). We show our results by applying the Drude model to direct reflectivity measurements (inverted triangles, with error bars) and to the Hill fit on our reflectivity datasets (continuous lines). Temperatures from Chau \cite{chau11} have been corrected transposing the correction made by Millot \cite{millot18} on another dataset of the same author \cite{chau01}.}
\label{fig:sigma_T}
\end{figure}

In Figure \ref{fig:sigma_T} we show the temperature-dependent DC conductivity of water and C:H:N:O mixture. After a rapid arise from $4000$ to $10000$ K, conductivity quasi-saturates, following the reflectivity behaviour.
In this region, at $15000$ K ($\simeq 2$ Mbar), conductivity values are $\sim 2.1$ and $\sim 4.1 \cdot 10^2$ S/cm for water and C:H:N:O, respectively. We notice that the C:H:N:O mixture conductivity is higher than water. A different behaviour was observed in multiple shock experiments. As already underlined, in this conditions the electronic contribution dominates over the ionic one. This situation is different from previous experiments \cite{nellis97, chau11} where the main contribution was ionic.\\
These data highlight that the use of water as a representative of planetary ices can be a too simplified picture for dynamo modelisation.\\

\subsection*{Refractive index}

The combined use of the reflectivity at two different frequencies also allows us to infere a measurement of the complex refractive index of the shocked sample. From the couple $(n_b, \sigma_{dc})$, extracted as explained in Subsection \ref{sub:cond}, we can compute the complex refractive index along the Hugoniot:

\begin{subequations}
\begin{align}
\tilde{n}(\omega) &= \left( n_b^2 + i \frac{\sigma_{dc}}{\epsilon_0 \omega (1 - i \omega \tau)} \right)^{1/2}  \\
\tilde{n}(2\omega) &=  \left( n_b^2 + i \frac{\sigma_{dc}}{2 \epsilon_0 \omega (1 - 2 i \omega \tau)}  \right)^{1/2}.
\end{align}
\end{subequations}

\begin{figure}[h!]
\centering
\includegraphics[width=\textwidth]{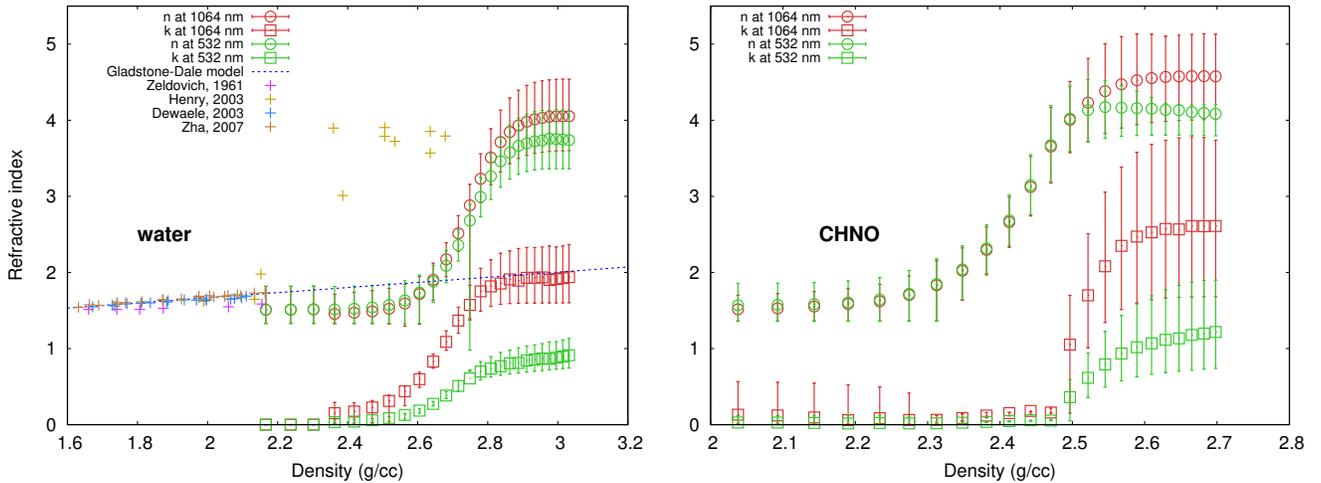}
\caption{\textbf{Left.} Real ($n$) and imaginary ($k$) part of the shock-compressed water refractive index \textit{vs} density. The dashed line is a Gladstone-Dale model for $n$ shown in literature \cite{batani15}. \textbf{Right.} Real and imaginary part of the shock-compressed C:H:N:O mixture refractive index \textit{vs} density.}
\label{fig:refractive_index}
\end{figure}

The real and imaginary part of the refractive indices of water and C:H:N:O mixture at both laser frequencies are shown in Figure \ref{fig:refractive_index}. Low-density real refractive index values are comparable to the results in the literature \cite{zeldovich61, henry03, dewaele03, zha07}. At densities around $2.6$ g/cm$^3$, the water real refractive index starts to increase from values comparable to those given by the Gladstone-Dale model \cite{batani15} to a saturation value of around $3.5 - 4$. This value is very similar to previous results \cite{batani15, henry03}, although they found it at around $2.4$ g/cm$^3$, in an opaque regime where reflectivity could not be measured. It has been recently noticed \cite{millot18} that this is in contrast with the Fresnel reflectivity estimation obtained with $n = 3.5$. Our data are not affected by this inconsistency, since we find $n \simeq 3.5$ at $2.8$ g/cm$^3$, where water reflectivity is $\sim 20\%$ .\\
The increase of the real and imaginary part of the refractive index are a marker of the transition to a metallic state. Our data show that this transition takes place between $2.5$ and $2.8$ g/cm$^3$. For C:H:N:O, the sudden increase in both the real and imaginary refractive indices takes place around $2.5$ g/cm$^3$.\\

%
%
%

\section*{Conclusions}

We studied the behaviour of shock-compressed water and C:H:N:O mixtures at extreme conditions in the Warm Dense Matter regime, reaching pressures up to $2.8$ Mbar and temperatures of $24 000$ K.\\
We obtained $\rho$-$p$-$E$ equation of state, temperature, optical reflectivity, and electronic contribution to the electrical conductivity of pure H$_2$O and C:H:N:O mixtures along their principal Hugoniot. We found that the only difference in the $\rho$-$p$ relations of water and C:H:N:O can be completely explained by the difference in the initial densities. Their $T$-$p$ relations are comparable, although possible small discrepancies could not be distinguished.\\
The similarity between the equations of state of water and C:H:N:O confirms the validity of the Linear Mixing Approximation at planetary conditions \cite{bethkenhagen17}. Moreover, the studied Hugoniot states are consistent with the existence of an atomic fluid above $5000 - 6000$ K as recently expected by first-principles calculations \cite{meyer15}.\\
The reflectivity behaviour of water and C:H:N:O mixture are different. The reflectivity onset for C:H:N:O is at $0.8 - 0.9$ Mbar ($R_{532 \ nm} = 2.9 - 5.8\%$), slightly lower than water which is found at $1.1 - 1.2$ Mbar ($R_{532 \ nm} = 2.4 - 4.8\%$). At $1.5$ Mbar, C:H:N:O reflectivity at $1064$ nm gets higher than at $532$ nm showing a strong metallic behaviour. The reflectivity saturation values are higher for C:H:N:O than for water ($29\%$ against $24\%$ at $532$ nm, $41\%$ against $34\%$ at $1064$ nm).\\
Using the dual reflectivity measurement, conductivity and complex refractive index of shocked water and C:H:N:O mixture are obtained through a modified Drude model. Saturation values for conductivities are $\sim 2600$ and $\sim 4000$ S/cm for water and C:H:N:O, respectively. Our results suggest that, in a mantle composed by C:H:N:O mixtures, planetary dynamo could be sustained differently than expected if water is assumed as unique component.\\
Future experimental work should consider high pressure off-Hugoniot states to enlarge the studied scenario and explore thermodynamic conditions more relevant to planetary interiors.\\

\begin{table}[h!]
\centering

    \begin{tabularx}{0.7\textwidth}{  XXXX  }
    \hline
    \textbf{shot \#} & $\rho$ (g/cm$^3$) & $p$ (Mbar) & $E - E_0$ (kJ/g)\\
    
\hline
 & & &     \\
\multicolumn{4}{c}{\textbf{Pure water}} \\
    GK-680 & 2.74 $\pm$ 0.18 & 1.23 $\pm$ 0.05  & 39.2 $\pm$ 2.7 \\
    L1-19  & 3.00 $\pm$ 0.15 & 2.73 $\pm$ 0.06 & 91.3 $\pm$ 3.7\\
    
 & & &     \\
\multicolumn{4}{c}{\textbf{C:H:O mixture}} \\
    GK-687 &  2.48 $\pm$ 0.15 & 1.42 $\pm$ 0.04 & 51.9 $\pm$ 2.7 \\
    GK-706 &  2.24 $\pm$ 0.20 & 0.74 $\pm$ 0.03 & 25.5 $\pm$ 2.0\\
    GK-725 &   2.65 $\pm$ 0.13 & 2.55 $\pm$ 0.06 & 96.8 $\pm$ 3.8\\
    GK-744 &   2.32 $\pm$ 0.14 & 1.16 $\pm$ 0.04 & 41.1 $\pm$ 2.6\\
    GK-753 &   2.38 $\pm$ 0.22  & 0.94 $\pm$ 0.04 & 33.5 $\pm$ 2.7\\
    L1-29 &  2.86 $\pm$ 0.19 & 2.67 $\pm$ 0.07 &104.8 $\pm$ 5.1\\
    L2-82 &  2.37 $\pm$ 0.16 & 2.79 $\pm$ 0.11 & 99.6 $\pm$ 6.8\\
    L2-86 &  2.31 $\pm$ 0.19 & 1.84 $\pm$ 0.08  & 64.8 $\pm$ 5.3\\
    
 & & &     \\
\multicolumn{4}{c}{\textbf{C:H:N:O mixture}} \\

    GK-694  & 2.56 $\pm$ 0.19 & 1.73 $\pm$ 0.06 & 63.6 $\pm$ 4.0\\
    GK-696 & 2.78 $\pm$ 0.19 &1.58 $\pm$ 0.05 & 60.4 $\pm$ 3.4\\
    GK-712   & 2.79 $\pm$ 0.23 & 2.60 $\pm$ 0.08 & 99.4 $\pm$ 5.9\\
    GK-715  & 2.14 $\pm$ 0.19 & 0.60 $\pm$ 0.03 & 19.5 $\pm$ 2.1\\
    GK-745 & 2.18 $\pm$ 0.19  &  0.67 $\pm$ 0.03 & 22.3 $\pm$ 2.1\\
    L2-102   & 2.65 $\pm$ 0.26 &1.57 $\pm$ 0.07 & 56.7 $\pm$ 4.8\\
    L2-130   & 2.44 $\pm$ 0.30 & 2.08 $\pm$ 0.10 & 74.4 $\pm$ 7.8\\
\hline

    \end{tabularx}
    
\caption{Experimental data on pure water, C:H:O mixture, and C:H:N:O mixture. The shot number  prefix GK, L1, and L2 identify the campaigns at GEKKO XII in January 2016, at LULI 2000 in February 2017, and at LULI 2000 in September 2017, respectively.}

\label{tbl:exp_data}
\end{table}


\bibliography{biblio_mixtures_scirep_titles}
\bibliographystyle{unsrt}
%

\section*{Acknowledgements}

We want to thank the GEKKO XII and LULI 2000 laser and support teams. We are grateful to Ronald Redmer, Martin French, and Mandy Bethkenhagen for the useful discussions. This research was supported by a French ANR grant to the POMPEI project (ANR-16-CE31-0008), the JSPS core-to-core program on International Alliance for Material Science in Extreme States with High Power Laser and XFEL, and the International Joint Research Promotion Program at the Osaka University. This work has taken advantage of the MECMATPLA international collaboration.

\section*{Author contributions statement}

T.O., A.B.-M., R.K., N.O., and A.R.. conceived the project. M.G., J.-A.H., T.O., P.B., A.B.-M., R.B., E.B., Y.F., F.L., K.M., N.O., Y.U., T.V., and A.R. conducted the experiments. M.G. analysed the results. M.G., J.-A.H, and A.R. wrote the paper. All authors reviewed the manuscript.

\section*{Additional information}

Supplementary infomation for this article is available. The authors declare no competing interests. The data that support the plots within this paper and other
findings of this study are available from the corresponding author upon
reasonable request.
%

\end{document}